\def\useieeelayout{0}
\def\showall{0}

\newcommand{\inConf}[1]{\if\showall1{\color{green!50!black}In Journal: #1}\else\if\useieeelayout1{#1}\fi\fi}

\newcommand{\inArxiv}[1]{\if\showall1{\color{blue}In ArXiV: #1}\else\if\useieeelayout0{#1}\fi\fi}

\if\useieeelayout1
\documentclass[letterpaper, 10 pt, conference]{ieeeconf}  
\IEEEoverridecommandlockouts                              %
\usepackage{graphicx}
\usepackage{amsmath}
\usepackage{optidef}
 
\usepackage{amsthm}
\usepackage{amssymb}
\usepackage{amsfonts}       
\usepackage{multicol}
\usepackage{cuted}
\usepackage{multirow}
\usepackage{caption, subcaption}
\newtheorem{theorem}{Theorem}[section]
\newtheorem{lemma}[theorem]{Lemma}
\newtheorem{proposition}[theorem]{Proposition}
\newtheorem{definition}{Definition}[section]
\newtheorem{problem}{Problem}[section]
\newtheorem{assumption}{Assumption}[section]
\newtheorem{remark}{Remark}[section]
\newtheorem{question}{Question}[section]
\newtheorem{conjecture}[theorem]{Conjecture}
\newtheorem{corollary}{Corollary}[theorem]
\usepackage{soul}
\usepackage{cite}
\usepackage[colorlinks=true,allcolors=blue]{hyperref}
\usepackage{comment}
\usepackage{algorithm}
\usepackage{algorithmic}
\usepackage{scalerel}
\usepackage{bm}
\usepackage{multirow}
\usepackage{rotating}
\usepackage{pifont}

\usepackage{textcomp}
\usepackage{multirow}
\usepackage{orcidlink}
\usepackage[intoc, english]{nomencl}
\makenomenclature

\title{\LARGE \bf
Resilient Substation Design for 500-Year Storm Events – Current State–of-the-Art and Challenges for Floodplain Management and Infrastructure Hardening

}

\author{Chinmay Shetty~\orcidlink{0009-0008-1656-3437} and Soham Ghosh~\orcidlink{0000-0002-6151-8183}
\thanks{Chinmay Shetty and Soham Ghosh are with Black \& Veatch, Overland Park, KS 66211, USA. Correspondence email (Soham Ghosh): sghosh27@ieee.org}%
}

\else
\documentclass[10pt]{article}
\usepackage{graphicx}
\usepackage{amsmath}
\usepackage{optidef}
 
\usepackage{amsthm}
\usepackage{amssymb}
\usepackage{amsfonts}       
\usepackage{multicol}
\usepackage{cuted}
\usepackage{multirow}
\usepackage{caption, subcaption}

\usepackage{xcolor}

\usepackage{cite}
\usepackage[colorlinks=true,allcolors=steelblue]{hyperref}
\usepackage{comment}
\usepackage{algorithm}
\usepackage{algorithmic}
\usepackage{scalerel}
\usepackage{bm}
\usepackage{url}
\definecolor{steelblue}{RGB}{70,130,180}

\usepackage[preprint]{tmlr}
\usepackage[intoc, english]{nomencl}
\makenomenclature

\title{The Art of of Resilient Substation Design for 500-Year Storm Events – Current State–of-the-Art and Challenges for Floodplain Management and Infrastructure Hardening}

\author{\name Chinmay Shetty \email shettyc@bv.com \\
      \addr Civil and Site Engineering Division \\ Black and Veatch KS 66211, USA\\
      ORCID: 0009-0008-1656-3437\\
      \AND
      \name Soham Ghosh \email Correspondence email: sghosh27@ieee.org \\
      \addr Power Delivery Division\\ Black and Veatch KS 66211, USA\\
      ORCID: 0000-0002-6151-8183
      }

\def\month{MM}  
\def\year{YYYY} 
\def\openreview{\url{https://openreview.net/forum?id=XXXX}} 

\fi

\begin{document}

\maketitle
\thispagestyle{empty}
\pagestyle{empty}

\begin{abstract}

This manuscript develops a unified, applications-oriented engineering framework for Substations of the Future that can withstand non-stationary 500-year flood events, addressing a critical gap in how erosion control, geotechnical stabilization, green infrastructure, and phased retrofit planning are currently treated in isolation. Electrical substations are among the most flood-exposed nodes in the bulk power system, with tens of thousands of U.S. assets located in 100 and 500-year FEMA floodplains, leading to cascading outages and large annualized economic losses under climate-amplified storms. In response, the paper consolidates several standalone engineering design essentials into a single multi-scalar resilience framework beginning with articulating concrete block (ACB) revetments for flexible, permeable erosion protection that support aquifer recharge and LEED-oriented heat-island mitigation. Lime, cement, and fly-ash-based soil stabilization is explored along with green infrastructure, including permeable hardscape, bioswales, and targeted floodplain preservation. A phased roadmap for resiliency upgrades for existing brownfield legacy substations through deployable barriers and pumps, perimeter ACB and drainage upgrades, yard re-grading and equipment elevation is also presented. Collectively, these elements operationalize resilience against 0.2 percent annual exceedance probability events while reducing life-cycle costs and delivering co-benefits in stormwater quality, habitat restoration, and long-term grid reliability.

\end{abstract}
\textit{\textbf{Keywords:}} Resilient substation design, multi-scalar resilience, lime stabilization, flood resiliency.


\nomenclature{ACB}{Articulating Concrete Block}
\nomenclature{AEP}{Annual Exceedance Probability}
\nomenclature{ASCE}{American Society of Civil Engineers}
\nomenclature{BFE}{Base Flood Elevation}
\nomenclature{CAH}{Calcium Aluminate Hydrate}
\nomenclature{CTB}{Cement Treated Base}
\nomenclature{CSS}{Cement Stabilized Subgrade}
\nomenclature{CSH}{Calcium Silicate Hydrate}
\nomenclature{DFIRM}{Digital Flood Insurance Rate Map}
\nomenclature{DOE}{U.S. Department of Energy}
\nomenclature{EPA}{U.S. Environmental Protection Agency}
\nomenclature{FEMA}{Federal Emergency Management Agency}
\nomenclature{FIRM}{Flood Insurance Rate Map}
\nomenclature{GI}{Green Infrastructure}
\nomenclature{GIS}{Geographic Information System}
\nomenclature{HAZUS}{Hazards-United States}
\nomenclature{HEC-FIA}{Hydrological Engineering Center's Flood Impact Analysis}

\nomenclature{HIFLD}{Homeland Infrastructure Foundation Level Data}
\nomenclature{HRC-RAS}{Hydrologic Engineering Center – River Analysis System}
\nomenclature{IDF}{Intensity–Duration–Frequency (curve)}
\nomenclature{IPCC}{Intergovernmental Panel on Climate Change}
\nomenclature{LEED}{Leadership in Energy and Environmental Design}
\nomenclature{NERC}{North American Electric Reliability Corporation}
\nomenclature{NERC CIP-014}{NERC Critical Infrastructure Protection Standard CIP 014 (Physical Security)}
\nomenclature{NOAA}{National Oceanic and Atmospheric Administration}
\nomenclature{QER}{Quadrennial Energy Review}
\nomenclature{SCADA}{Supervisory Control and Data Acquisition}
\nomenclature{SETS}{Social–Ecological–Technological Systems}
\nomenclature{UCS}{Unconfined Compressive Strength}
\nomenclature{USACE}{U.S. Army Corps of Engineers}
\nomenclature{SHAP}{Shapley Additive Explanations}
\nomenclature{LIME}{Local Interpretable Model-Agnostic Explanations}
\nomenclature{SRI}{Solar Reflectance Index}
\nomenclature{CLASIC}{Community-enabled Lifecycle Analysis of Stormwater Infrastructure Costs}

\printnomenclature

\section{Introduction}\label{sec_intro}

Electrical substations are critical nodes in the power grid, yet they are increasingly vulnerable to climate-amplified hydrometeorological extremes. Traditional stationary design storms, such as those defined in NOAA Atlas 14 for the 100-year return period, now underestimate observed and projected extremes by approximately 20–50\% because of anthropogenic warming and associated non-stationarity in precipitation regimes. Recent national-scale analyses indicate that about 9–14\% of U.S. transmission and distribution, predominantly substation assets, are located within FEMA 100-year flood zones, and a further 10\%  within the FEMA mapped 500-year zones. Estimates and ranges vary depending on the level of granularity and the timeline of the survey being considered. An important validation in this area comes from the recently published \cite{tufail2026assessing}, where the authors combine FEMA Special Flood Hazard Areas (SHFA) with NOAA's coastal composite (covering storm surges, high-tide flooding, sea-level rise) to develop a unified view of vulnerability of US energy infrastructure. The authors found that of the 21,988 energy infrastructures studied, 9.2\% (2,021 energy infrastructures) fall within FEMA SHFA, and another 9.5\% (2,078 energy infrastructures) fall under the NOAA coastal composite hazard zone. A joint composite map of these two sources reveals that 14.4\% (3,174 energy infrastructures) fall in the flood hazard zone. Another established study by \cite{ref1_qiang2019flood} highlighted that about 7.7 \% of U.S. critical infrastructures are located in the FEMA 100-year floodplain (referred to as baseline exposure), with a corresponding 14.5\% (919 out of 6,341 critical energy infrastructures) falling within the FEMA 100-year floodplain, and hence are susceptible to flooding and unplanned outages. \\
Historical events highlight the consequences of these exposures. During Hurricane Harvey (2017), multiple substations in Texas were inundated, while Hurricane Ida (2021) disrupted service to approximately 1.2 million customers, demonstrating how flood induced damage at a limited number of sites can cascade into widespread outages and systemic stress across regional grids. Forward-looking assessments suggest that, under continued warming scenarios, similar climate-driven events could generate average annual U.S. flood-related grid losses exceeding USD 750 million by 2100 when both localized (urban) and riverine flooding are considered \cite{ref3}. These observations point toward the need to move beyond stationary design assumptions and toward explicitly resilience-oriented substation planning and retrofit strategies.
Traditional mitigation approaches for flood-exposed substations include rigid concrete floodwalls, bulkheads, riprap armoring, and full platform elevation. While these measures can reduce inundation depth at specific sites, they are often capital-intensive, inflexible, and environmentally intrusive. In many cases, hard perimeter defenses increase surface runoff, diminish natural floodplain storage, and degrade local ecosystems, while providing limited capacity to adapt to evolving hazard profiles or compound events. Because most substations are legacy facilities that must remain in operation, utilities are further constrained by narrow outage windows and the need for retrofit solutions that can be implemented in phased, low downtime increments.
In practice, mitigation actions are frequently siloed. Erosion control, subgrade stabilization, and stormwater management are often designed independently, and flood modeling is not always fully integrated with site-scale grading, drainage, and equipment configurations. This fragmentation limits opportunities to exploit synergies between structural, geotechnical, and green infrastructure measures. A next-generation resilience framework must therefore treat these elements as interconnected components of a unified site development and risk management strategy.
\subsection{Motivation for the Work}
The motivation for developing a resilient substation framework arises from the convergence of intensifying climate hazards and the essential role of substations in sustaining modern economies and communities. Intensified storm events and changing flood probabilities now threaten the continuity of the electrical grid, which underpins everything from critical health and safety services to digital infrastructure. Events like Hurricanes Harvey and Ida, which flooded substation sites and left large populations without power for prolonged periods, demonstrate that localized failures at a small number of nodes can cascade into widespread outages and disproportionate impacts on vulnerable communities with limited adaptive capacity. \\
Projections of increasing flood-related damages over this century further emphasize the urgency of transitioning from reactive recovery to proactive, risk-informed design and retrofit.
With tens of thousands of substations located in 100 and 500-year floodplains, regulatory frameworks such as NERC CIP 014, ASCE 24 Class IV, and FEMA Risk MAP collectively point toward the necessity of enhanced physical security and flood resilience for critical nodes. Conventional solutions, including riprap armoring and wholesale elevation platforms, often entail high capital and opportunity costs and may not fully leverage the potential of integrated geotechnical and green infrastructure measures to reduce risk at lower life-cycle costs. Because most assets are existing facilities, utilities must pursue hybrid strategies that combine quickly deployable measures with long-term, sustainable upgrades, as evidenced by leading utilities that have begun to embed flood resilience into asset management and capital planning processes.

\subsection{Research Gap and Manuscript Contribution}
Electrical substations are among the most flood-exposed nodes in the bulk power system, yet engineering literature has not kept pace with the compounding risks these assets now face. It is estimated that approximately 9–14\% of U.S. substations lie within FEMA 100-year flood zones, with limited and localized highlights on FEMA 500-year flood zones such as \cite{thakali2017flood}. Additionally, dual-source flooding (FEMA SHFA + NOAA coastal composite), simultaneous riverine and pluvial inundation, can affect roughly 15\% of the national energy infrastructure fleet (predominantly being some form of substation) \cite{ref4_tufail2026assessing}. Despite this exposure, mitigation measures for erosion control, geotechnical stabilization, and stormwater management have largely been developed and applied in disciplinary isolation, with no unified design standard that coordinates them as components of a single resilience strategy \\
As shall be explored, there exists a three-dimensional research gap that is particularly consequential. 
\begin{enumerate}
    \item First, articulating concrete block (ACB) revetment research has concentrated on generic channel and levee applications \cite{ref5, ref6_delport2021incipient}; no coupled hydraulic-structural model exists for the substation-specific geometry of yard grading, cable trench exits, and equipment pads.
    \item Second, the extensive literature on lime and cement stabilization was developed almost entirely for roadway pavements under traffic loading; behavior under repeated inundation cycles, drawdown-induced effective-stress reversals, and freeze-thaw degradation in flood-exposed substation subgrades remains poorly characterized \cite{ref7_chou2024stabilization}.
    \item Third, green infrastructure guidance rarely quantifies the site-scale contributions of bioswales and permeable hardscape to substation resilience using integrated hydrologic-economic tools.
\end{enumerate}  
At the system and multi-hazard scale, additional voids persist. The authors \cite{ref8_mensah2016efficient} developed a Bayesian network framework for quantifying hurricane-induced power system outages, demonstrating that the compound effect of wind and storm surge on substations creates spatiotemporal loss patterns that single-hazard design criteria cannot capture. For seismic hazards, a resilience assessment framework for electrical substations is proposed that models bushing fracture, recovery scheduling, and functionality under earthquake loading \cite{ref9_liang2023seismic}. Extended fragility estimation to account for internal component configurations and short-circuit faults, both studies underscoring that seismic fragility of substation equipment is highly equipment-class-specific and cannot be approximated by generic residential or commercial fragility curves \cite{ref10_ahumada2025seismic}. Hurricane and storm-surge vulnerability further mapped substation outage probabilities under historical hurricane tracks to county-level community impact indices, revealing that coastal substation inundation accounts for a disproportionate share of post-event load loss. Comprehensive insights presented in this section are extended in Table \ref{tab:literature_summary_and contributions}. All these hazard domains show that intensity-duration-frequency curves used in infrastructure design must be updated for non-stationarity under climate change, yet no current substation resilience framework operationalizes non-stationary IDF estimates alongside site-scale engineering controls.
Taken together, the literature reveals the absence of a unified, standards-aligned "Substation of the Future" framework that explicitly couples erosion control, geotechnical stabilization, green infrastructure, and phased resilience planning against a non-stationary 500-year design standard.

Against this backdrop, the distinct contribution of this manuscript is that it advances the state of practice across three interconnected dimensions.

\begin{enumerate}
  \item It consolidates erosion control (ACB revetments), geotechnical stabilization (lime, cement, and fly ash), and green infrastructure into a single, unified design framework anchored to a 500-year (0.2\% AEP) non-stationary flood standard—an integration absent from the prior literature.
  \item It provides quantitative design tools grounded in applicable North American standards: ACB moment-equilibrium stability equations, soil strength benchmarks (UCS 100–600 psi; resilient modulus gains up to one order of magnitude), and green infrastructure sizing protocols aligned with EPA, USACE, and FEMA guidance.
  \item It introduces a five-phase retrofit roadmap (Phases 0–4) for energized legacy substations that delivers incremental progress toward full resilience with minimal service interruption, reducing projected lifecycle costs while delivering co-benefits, including LEED credits, aquifer recharge, and reduced urban heat island effects.
\end{enumerate}

\begin{table*}[t]
\small
\centering
\caption{Comparison of prominent related work in the field of substation resilience from a site engineering design standpoint.}
\label{tab:literature_summary_and contributions}
\renewcommand{\arraystretch}{1.2} 
\begin{tabular}{p{2cm} p{6cm} p{2cm} p{6cm} }

\hline
\textbf{Research Gap Area} &
\textbf{Prior Literature State}&
\textbf{Key References}&
\textbf{Manuscript's Contribution}
\\
\hline

Nationwide critical infrastructure exposure &
Previous studies had largely assessed flood exposure for general populations and economies, or for critical infrastructures only at city or regional scales, without a consistent national-level quantification. &
\cite{ref1_qiang2019flood, ref4_tufail2026assessing} & 
This manuscript provides a nationwide spatial assessment of 100‑year flood exposure for major U.S. critical infrastructure sectors by overlaying FEMA flood maps with the USGS National Structures Dataset. \\
\hline

Articulating Concrete Block (ACB) for substations &

ACB stability methods (factor-of-safety, moment-equilibrium) were developed and validated for generic channels, levees, and dam spillways; no equivalent methodology existed for substation-specific geometry or coupled yard-drainage configurations. &
\cite{ref11_delport2021incipient, ref5} &
Applies ACB shear- and velocity-based stability design (Equations 1–5) specifically to substation embankments, perimeter channels, and equipment pads; demonstrates lifecycle cost advantages relative to riprap and cast-in-place concrete in a substation context. \\
\hline

Geotechnical stabilization in flood-prone subgrades & 
Lime, cement, and fly-ash stabilization were well-characterized for roadway pavements; their performance under repeated substation floodplain inundation cycles, drawdown, and freeze-thaw degradation was not systematically addressed. &
\cite{ref13_diniz2024lime, ref14} &
Synthesizes lime, cement (CMS/CSS/CTB), and fly-ash stabilization performance for flood-exposed substations; provides a comparative technical assessment and establishes strength (100–600 psi UCS) and durability targets appropriate for floodplain subgrades. \\ 

\hline
Urban comprehensive impact quantification &
Prior studies mostly used HAZUS or HEC-FIA separately, focused on direct physical damage or limited case studies, and rarely integrated socio-economic losses, life-loss estimation, and depth–velocity-based vulnerability mapping at a multi-community scale in Iowa. &
\cite{ref2_alabbad2022comprehensive} & 
This manuscript develops and applies an integrated, GIS-based framework combining HAZUS, HEC-FIA, and depth–velocity floodplain vulnerability curves to quantify multi-dimensional economic losses, potential fatalities, and hazard classes for several Iowa communities  \\
\hline

Dual-flood exposure & 
Prior national-scale studies assessed U.S. energy infrastructure flood risk using a single hazard source (typically FEMA riverine SFHAs or individual coastal products). & 
\cite{ref4_tufail2026assessing, ref15} & 
This study integrates FEMA Special Flood Hazard Areas and NOAA coastal flood composites into a unified national exposure assessment for nearly 22,000 U.S. energy facilities, quantifying sector- and subtype-specific hotspot regions \\
\hline

Multi-scalar resilience under compound hazards &
Single-hazard, single-return-period design-storm criteria dominated practice; compound flooding, cascading grid failures, and multi-hazard interactions (pluvial + riverine + surge) at the system and portfolio scale were not integrated into a substation resilience standard. &
\cite{ref16_prime2018protecting, ref17, ref18_markolf2021re, ref19_kim2022leveraging} &
Presents a three-scale resilience framework that explicitly links non-stationary IDF design to safe-to-fail yard configurations, hybrid gray-green controls, and portfolio-level risk-based prioritization across substation fleets.\\
\hline
Hurricane and storm-surge impact on substations &
Dueñas-Osorio characterized hurricane and storm-surge induced outage patterns using fragility curves and probabilistic models, but failed to offer site-level hardening specifications for reducing that vulnerability. &
\cite{ref8_mensah2016efficient} &
The manuscript addresses hurricane and surge exposure through perimeter ACB hardening, deployable flood barriers, elevating equipment above surge design levels, and GIS-enabled microgrid islanding measures designed to reduce the quantified fragility. \\
\hline

\end{tabular}
\end{table*}

The remainder of the manuscript is organized as follows. Section \S\ref{sec_floodplain types} sets the stage for the readers by providing a high-level overview of the types of floodplains designated by FEMA and addresses common misconceptions. Section \S\ref{sec_ACB} presents articulating concrete blocks (ACBs) based resilience improvements for erosion control and outlines installations for tapezoidal channel lining, embankment and creek cross, and typical cross-sections. Section \S\ref{sec_GI} presents a comprehensive evaluation of resiliency improvements through green infrastructure methods; green infrastructure enhances flood resilience by managing stormwater at its source and preserving floodplains, thereby reducing runoff, peak flows, and flood risks under increasingly intense precipitation conditions. Section \S\ref{sec_soil_stabilization} covers various soil stabilization methods for flood-prone subgrades, including lime stabilization, cement stabilization, and fly ash stabilization, along with a comparative assessment between these methodologies. Section \S\ref{sec_rexisting_stations} provides a phased upgrade framework for existing substations, beginning with inventory setup and immediate protection to long-term full system resilience. Finally, \S\ref{sec_conclusion} concludes the article and provides future research direction.  
\section{Types of Floodplains Designated by FEMA}
\label{sec_floodplain types}
\begin{figure*}[htbp]
    \centering

    \subfloat[Floodplains clearly defined by natural features where 100-year and 500-year floodplains are quite similar in extent. \label{fig:Floodplain_1}]{%
        \includegraphics[width=0.80\linewidth]{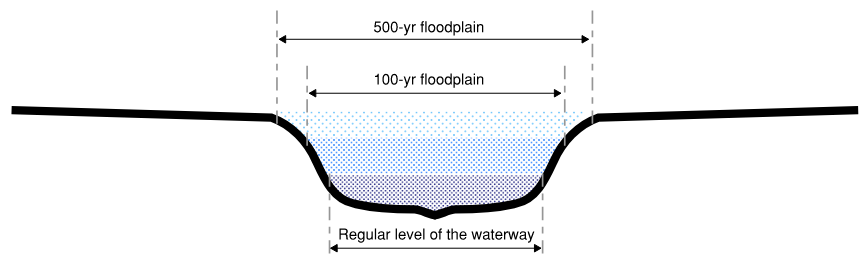}
    }
    \hfill

    \subfloat[Due to the relatively small elevation difference, 100-year and 500-year floodplains coverage are very different. \label{fig:Floodplain_2}]{%
        \includegraphics[width=0.80\linewidth]{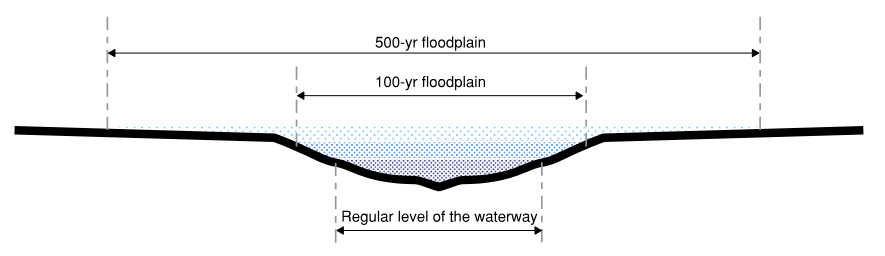}
    }
    \hfill

    \caption{Variance in 100-year and 500-year floodplain due to topography.}

    \label{fig:phased_approach}
    
\end{figure*}

The Federal Emergency Management Agency (FEMA) defines a floodplain as any land area susceptible to being inundated by floodwaters from any source. In practice, FEMA operationalizes “floodplain” through its Flood Insurance Rate Maps (FIRMs), which delineate flood hazard zones. The most commonly referenced is the 1‑percent annual chance floodplain (the “100‑year floodplain” or Special Flood Hazard Area), but FEMA maps also show 0.2‑percent annual chance (500‑year) floodplains and areas of minimal or undetermined risk. Many FEMA- and state-derived explanations emphasize that high‑risk floodplains are often relatively flat land adjacent to rivers, streams, lakes, or coastal areas that can be inundated during periods of high water. These areas function as natural storage and conveyance for floodwaters during storm events, which is why land‑use controls and building standards often focus on them. More formally, the types of floodplains designated by FEMA are: 
\begin{itemize}
    \item The \textbf{1 percent (100-year) floodplain} is the land that is covered in water during a flood event that has a 1 percent chance of being equaled or exceeded each year. 
    \item The \textbf{0.2 percent (500-year) floodplain} is the land that is covered in water during a flood event that has a 0.2 percent chance of being equaled or exceeded each year. 
\end{itemize}
What must be emphasized here is that a common misconception exists as to the definition of a 100-year flood plain. A 100-year floodplain doesn't get inundated once every 100 years, but has a 1\% chance of being flooded every year. This means that, statistically speaking, the 1\% flood event has approximately a 26\% chance of occurring during a 30-year time horizon. It should also be emphasized that, depending on the topography of the land, such as in the vicinity of a river or a river with levees holding back the floodwater, a 100-year versus a 500-year flood plain may be quite similar in extent, see Figure \ref{fig:Floodplain_1}. On the contrary, for bayou or marshy areas with flat adjacent landscape with relatively small elevation difference, water may spread more broadly as it rises out of the bayou or marshland. In such areas, a 100-year versus a 500-year flood plain looks vastly different. This is illustrated in Figure \ref{fig:Floodplain_2}. 

\section{Resiliency Improvements through Articulating Concrete Block}
\label{sec_ACB}
\begin{figure*}[htbp]
    \centering

    \subfloat[\label{fig1a}]{%
        \includegraphics[width=1.0\linewidth]{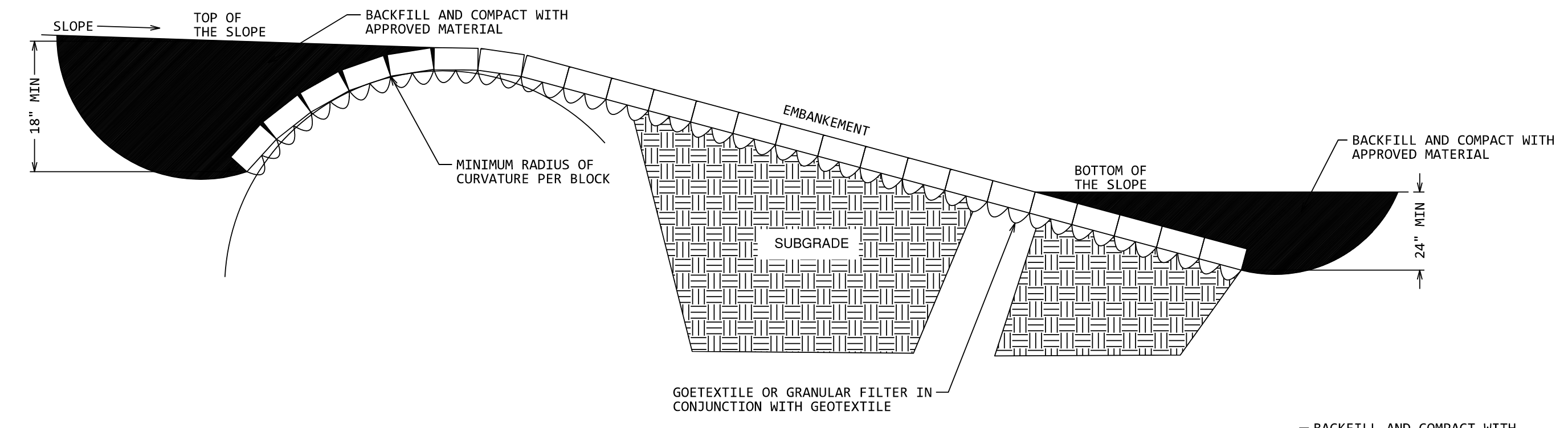}
    }
    \hfill

    \subfloat[\label{fig1b}]{%
        \includegraphics[width=1.0\linewidth]{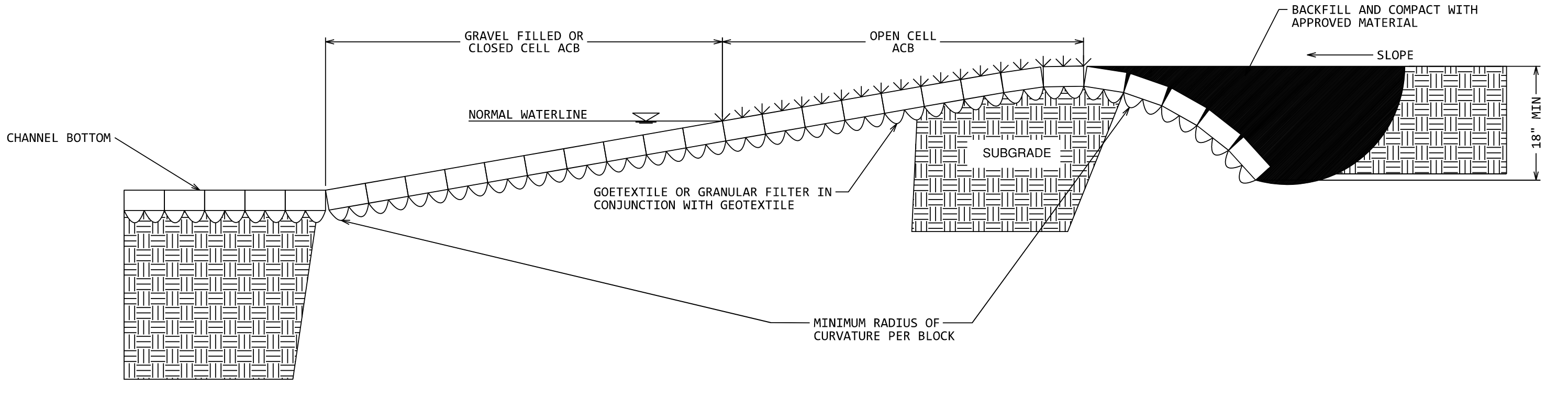}
    }
    \hfill
    
     \subfloat[\label{fig1c}]{%
        \includegraphics[width=1.0\linewidth]{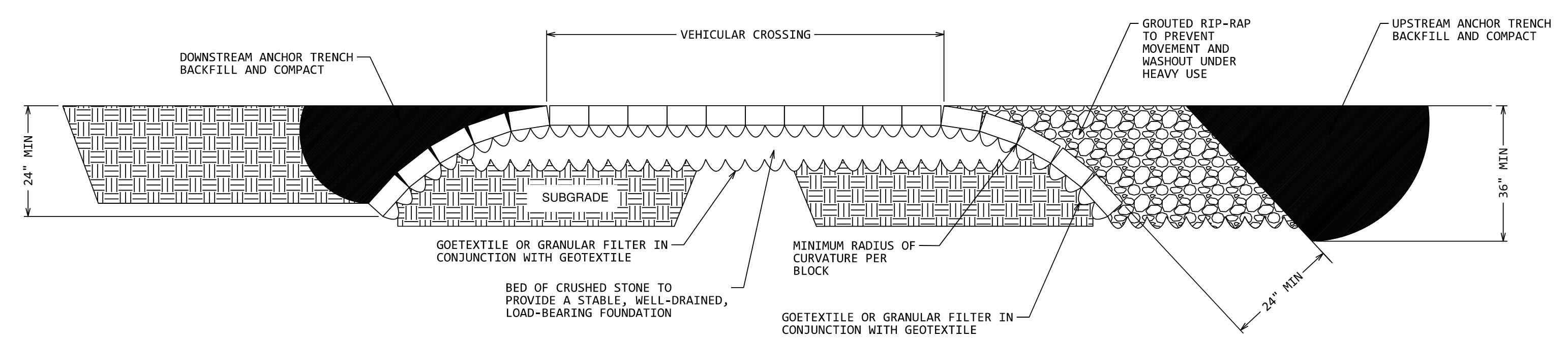}
    }

    \caption{ACB installation for trapezoidal channel lining, embankment and creek cross, typical cross-sections}
    \label{fig1}
\end{figure*}

Articulating concrete block (ACB) systems are widely used as engineered revetments for erosion control along channels, embankments, spillways, shorelines, and critical infrastructure platforms such as substations. An ACB system consists of a matrix of individual concrete blocks placed together to form an erosion-resistant, hydraulically stable overlay on a prepared subgrade, see Figures \ref{fig1a} - \ref{fig1c}. This matrix is installed over a filter layer that permits infiltration and exfiltration while retaining the underlying soil, typically using geotextiles, graded aggregates, or a combination of both. The blocks are designed to be dense and durable, while the overall system remains flexible and porous so that it can accommodate hydraulic loading, minor subgrade movements, and thermal expansion without loss of integrity \cite{ref20}. \\
The term articulating refers to the ability of individual blocks to conform to localized subgrade deformation while remaining interlocked or otherwise restrained. Restraint is provided by geometric interlock and, where used, by high-strength cables, ropes, geotextiles, or geogrids that tie the units into monolithic or semi-monolithic mats. ACBs are installed either as hand-placed units or as pre-assembled cabled mats placed with light equipment over the filter and prepared subgrade. Typical applications include drainage channels, riverbanks, bridge abutments and piers, dam and levee linings, spillways, boat ramps, low water crossings, retention and detention basins, lake and pond shorelines, and protection of buried pipelines and utilities. \\

From an environmental standpoint, ACB systems offer several advantages compared with conventional hard armor, such as riprap, gabions, soil cement, or cast-in-place concrete. Many block geometries incorporate vertical cores and open spaces that can be filled with soil and planted, allowing vegetation to establish through the revetment surface. With appropriate plant selection, vegetation can cover much of the exposed concrete, enabling the system to blend with the surrounding landscape while contributing to nutrient uptake and pollutant attenuation in stormwater. During high-intensity events, the underlying ACB layer resists scour even if the vegetation is overtopped or partially damaged, maintaining protection of the subgrade \cite{ref20}.
Because ACB revetments are permeable when installed over a properly designed filter, they help preserve natural drainage and treatment pathways, reduce runoff and peak flows, improve water quality, and promote aquifer recharge while preventing soil loss. These characteristics support their use in sustainable site design and green infrastructure applications. Within green building frameworks such as LEED, concrete hardscape products, including ACBs, can contribute to credits by reducing impervious cover, supporting habitat restoration, and utilizing local or recycled materials \cite{ref21}. Studies of permeable pavements installed over granular drainage layers, which are hydraulically analogous to vegetated ACB systems, have shown enhanced aerobic biodegradation of hydrocarbons and reductions in nutrient loads, reinforcing the water quality benefits of such assemblies. \\
Design of ACB revetments is commonly based on a factor of safety methodology that balances hydraulic driving forces against the resisting capacity of the block system \cite{ref22}. For steady, approximately uniform flow, the boundary shear stress is used for design as expressed in Equation \ref{eqn_1}.
\begin{equation}
\label{eqn_1}
\tau_{des} = \gamma RS_f \
\end{equation}
where $\tau_{des}$ is the unit weight of water, $R$ is the hydraulic radius, and $S_f$ is the energy slope, often taken as the channel bed slope. Design discharge and mean velocity are obtained using Manning’s equation, see in Equation \ref{eqn_2},
\begin{equation}
\label{eqn_2}
Q = V_{des} A = \frac{1}{n} AR^{2/3}S_f^{1/2} \
\end{equation}
where $Q$ is discharge, $A$ is flow area, and $n$ is Manning’s roughness coefficient. Each ACB product is characterized by critical hydraulic parameters, typically a critical shear stress $\tau_c$ and a corresponding critical velocity $V_c$, determined from flume and overtopping tests \cite{ref22}. \\
A shear-based factor of safety is defined as in Equation \ref{eqn_3}.
\begin{equation}
\label{eqn_3}
SF_{\tau} = \frac{\tau_c}{\tau_{des}} \
\end{equation}
and design is considered acceptable when this factor meets or exceeds a specified target, usually between 1.3 and 2.0, depending on project importance and uncertainty. The core stability check is a moment equilibrium analysis carried out on a representative block at incipient rotation about its downstream edge. The factor of safety with respect to overturning is written per Equation \ref{eqn_4}.
\begin{equation}
\label{eqn_4}
SF_{M} = \frac{\sum M_{resisting}}{\sum M_{overturning}}\
\end{equation}
where resisting moments arise from submerged block weight, normal forces, and interblock friction, while overturning moments are generated by hydrodynamic drag and lift acting on the block face \cite{ref22}. \\
Hydrodynamic forces are typically represented by Equation \ref{eqn_5}.
\begin{equation}
\label{eqn_5}
F_D = C_D \frac{1}{2} \rho V^2_{des} A_f, F_L = C_L\frac{1}{2}\rho V^2_{des} A_f\
\end{equation}
where $C_D$ and $C_L$ are drag and lift coefficients, $\rho$ is water density, and $A_f$ is the projected flow facing area of the block. Design manuals provide closed-form expressions and tabulated examples for calculating $SF_M$ by combining these forces with block geometry and moment arms. Additional checks are recommended for loss of intimate contact between the mat and subgrade and for localized edge instability at downstream boundaries, with analogous factors of safety defined for each mode. Specifications also prescribe detailing such as minimum cover lengths, toe embedment, allowable block protrusion, subgrade smoothness, and filter design to ensure that computed safety factors are realized in practice. \\
Recent research extends these methods through a shear and velocity-based stability assessment for channelized and overtopping flow conditions. In this approach, drag and lift in the moment balance are expressed explicitly as functions of both boundary shear and local velocity, calibrated against large databases of ACB performance tests. This combined treatment of $\tau_{des}$ and $V_{des}$ has been shown to improve prediction of both stable and failure states relative to shear only formulations \cite{ref23_cox2014articulated}. \\

Proper installation is critical to achieving the intended hydraulic and structural performance of ACB revetments. The subgrade must be graded to the design profile and provide a firm, unyielding foundation, free of protrusions or debris that could cause individual units to project significantly above adjacent blocks. A filter layer is then placed to retain the subgrade soil while permitting bidirectional flow, after which the ACB units or mats are installed. Open cells or joints are typically filled with granular material or soil, and vegetation is established where desired using seeding or mulching techniques. Cabled ACB systems assemble blocks into mattresses that can be placed rapidly with cranes, including on steep or underwater slopes, while non-cabled systems rely on the interlocking geometry of individually placed blocks and are advantageous where access is limited or plan geometry is complex. \\
In terms of economics, ACB systems are consistently presented as cost-effective hard armor solutions relative to riprap, gabion mattresses, and cast-in-place concrete for comparable levels of hydraulic protection. Cost advantages arise from reduced armor thickness, lower excavation and transport requirements, modular installation, and relatively low maintenance over the design life \cite{ref24}. Life cycle cost analysis frameworks, commonly used for concrete and block systems, can be applied to ACBs and competing revetments to compare net present costs over multi-decadal time horizons, capturing both initial construction and recurring maintenance and repair. Case-based evidence from channel restoration and flood protection projects indicates that, while initial unit costs per area for ACB and riprap can be similar, differences in required thickness, hauling volumes, and post-event repair frequency often lead to lower total life cycle cost for ACBs, especially where access is difficult or outages are expensive. These performance and economic characteristics make ACB revetments a strong candidate for climate-exposed infrastructure where durability, resilience, and long-term cost control are key objectives. \\


\section{Resiliency Improvements through Green Infrastructure} \label{sec_GI}
\begin{figure*}[htbp]
    \centering

    \subfloat[Green infrastructure improvements through bioswale. \label{fig:green_a}]{%
        \includegraphics[width=0.650\linewidth]{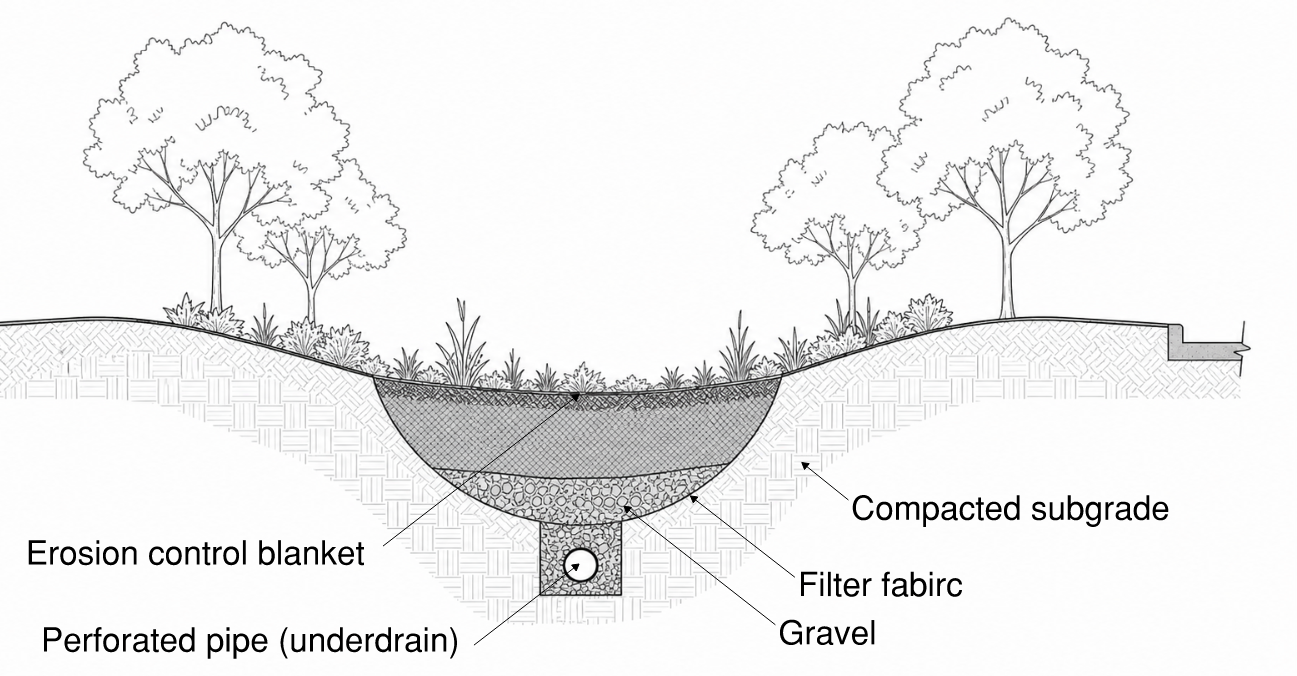}
    }
    \hfill

    \subfloat[Green infrastructure improvements through porous pavements. \label{fig:green_b}]{%
        \includegraphics[width=0.650\linewidth]{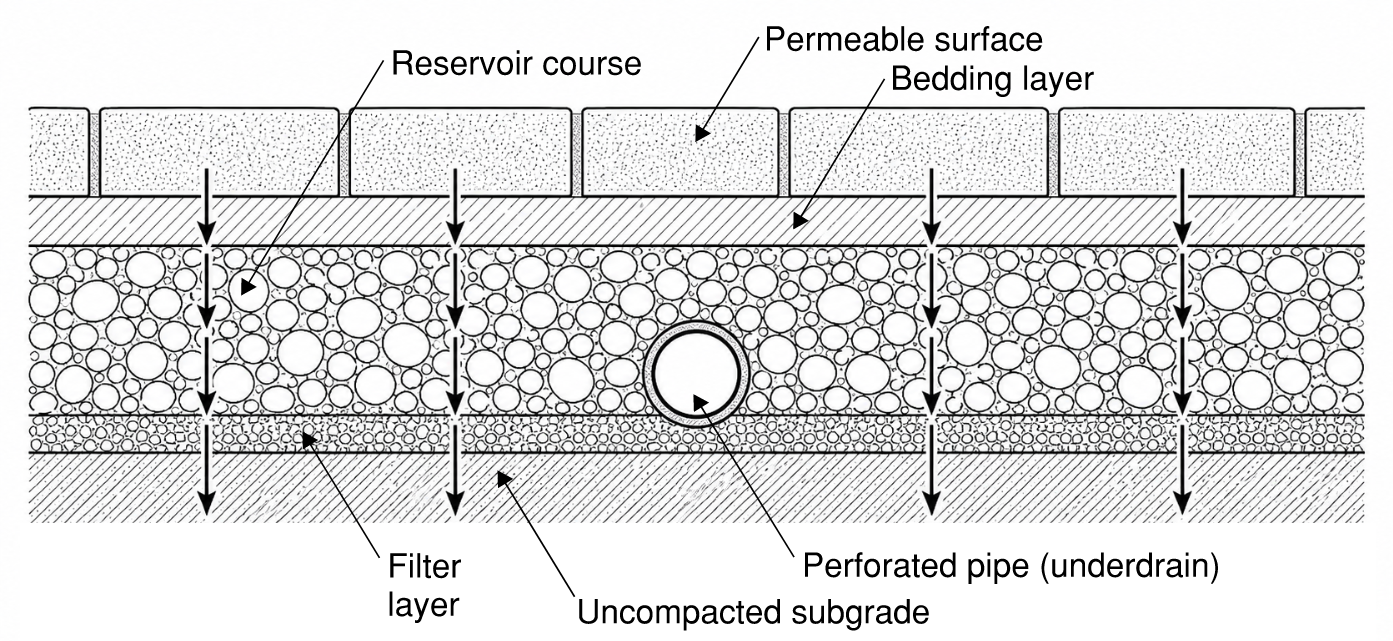}
    }
    \hfill
    
     \subfloat[Green infrastructure improvements through native vegetation restoration. \label{fig:green_c}]{%
        \includegraphics[width=0.650\linewidth]{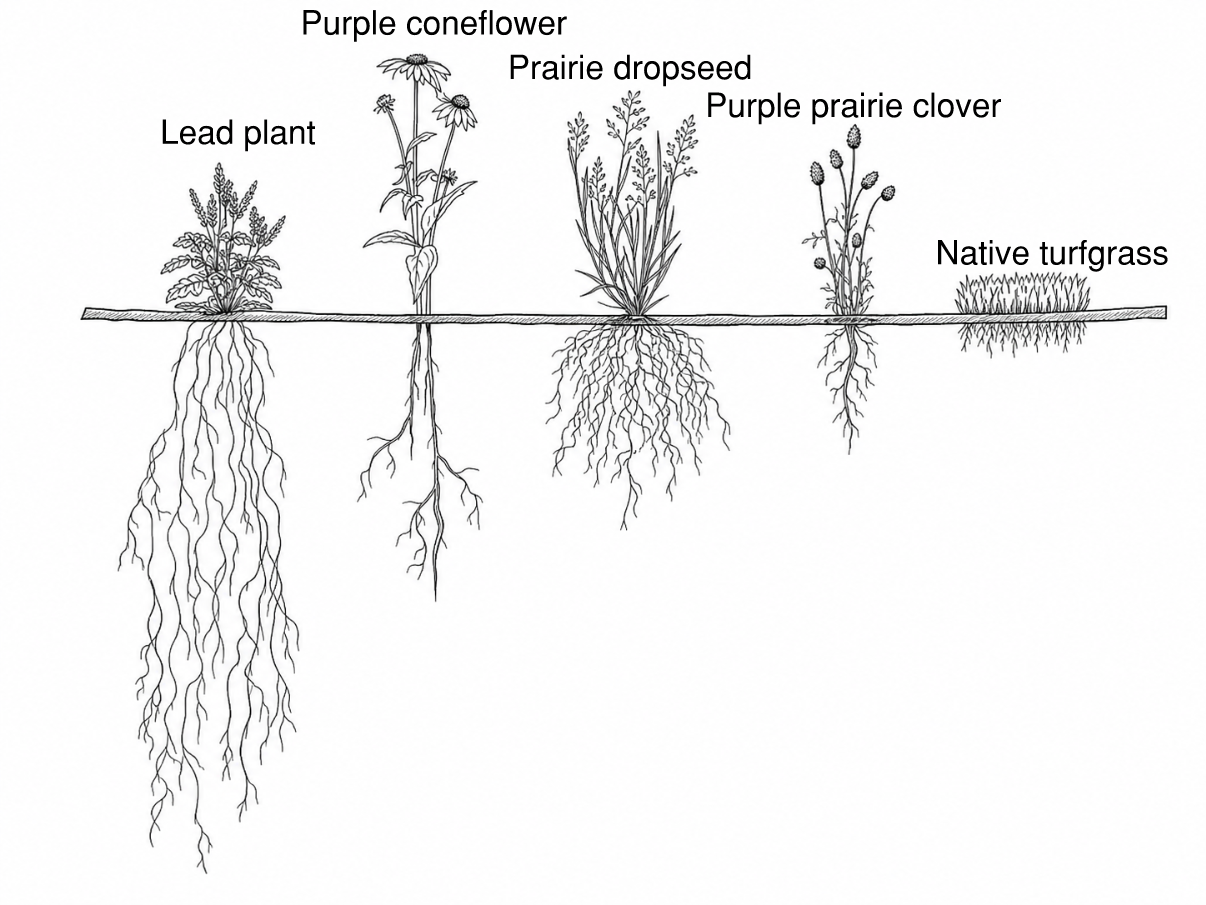}
    }
    \hfill

    \caption{Components of Milwaukee's Greenseams® program helping prevent future flooding and water pollution.}
    \label{fig:hybrid_energy_sources}
\end{figure*}

Green infrastructure plays a critical role in mitigating flood risk by slowing, storing, and infiltrating stormwater while preserving the natural functions of floodplains. As heavy precipitation events become more frequent and intense with a warming climate, both localized and riverine flooding are expected to increase across many parts of the United States, which will elevate annual flood damages and disproportionately affect communities with limited adaptive capacity. In this context, integrating green infrastructure into stormwater and floodplain management strategies is an essential component of community resilience to current and future flood hazards \cite{ref3}. \\
Green infrastructure for localized flooding focuses on managing stormwater at or near its source. Localized flooding typically occurs when rainfall exceeds the capacity of urban drainage and sewer systems, leading to ponding in streets, basements, and other low-lying areas. Practices such as permeable pavements, rain gardens, and bioswales enhance infiltration capacity, reduce runoff volumes, and attenuate peak flows, thereby lowering the likelihood that stormwater will overwhelm pipe networks. Detention and retention systems, including ponds and constructed basins, add further volume control and can be combined with vegetated practices to improve water quality through sedimentation, filtration, and biogeochemical uptake of pollutants. Hydrologic and hydraulic modeling tools enable communities to assess how different combinations of green and gray infrastructure affect runoff, peak flows, and water quality under current and projected precipitation regimes, and support the selection and sizing of green infrastructure measures that meet flood reduction and regulatory objectives within spatial and budgetary constraints [3].\\
For riverine flooding, which occurs when flows exceed the conveyance capacity of river channels and inundate adjacent floodplains, green infrastructure is most effective when combined with floodplain preservation and restoration. Conserving land within or adjacent to floodplains maintains or enhances the ability of these areas to temporarily store floodwaters, slow flow velocities, and reduce downstream peak discharges, thereby complementing structural measures such as levees and floodwalls. These strategies also reduce damage to critical infrastructure and property by keeping development out of high-risk zones and maintaining the ecological functions of riparian corridors \cite{ref25}. Geographic information systems (GIS) and hydraulic modeling are central to prioritizing where green infrastructure and land conservation will yield the greatest flood mitigation benefit per unit cost. GIS can identify flood-prone areas, characterize soil infiltration capacity, map development pressure, and screen parcels that are both hydrologically valuable and economically feasible to conserve, while models quantify stormwater volumes captured, stored, and infiltrated under different scenarios and estimate associated reductions in flood depths and damages \cite{ref26}.\\
Several decision support tools have been developed to operationalize these analyses. The U.S. Environmental Protection Agency’s National Stormwater Calculator allows users to evaluate site-scale hydrology, including future climate conditions, and estimate the performance and costs of a range of green infrastructure practices. The Community-enabled Lifecycle Analysis of Stormwater Infrastructure Costs (CLASIC) tool integrates GIS and scenario analysis to assess future precipitation, runoff reduction, costs, and co-benefits such as water quality improvements and ecosystem services across different stormwater management strategies. Complementary guidance documents from federal agencies describe how green infrastructure can be funded as a climate-resilient hazard mitigation activity, particularly where it reduces localized flooding, stream erosion, and flood damage \cite{ref25}. For coastal and fluvial settings, agencies have also highlighted the role of natural and nature-based solutions such as wetlands, dunes, and floodplain restoration in reducing storm surge impacts and fluvial flood risk. \\
A notable example of floodplain-oriented green infrastructure is the Greenseams program in the Milwaukee metropolitan area. In partnership with The Conservation Fund, the Milwaukee Metropolitan Sewerage District has strategically acquired and protected approximately 6,000 acres of flood-prone land characterized by water-absorbing soils and high development pressure. By preserving these landscapes in their natural or semi-natural state through a combination of green infrastructure methodologies (see Figure \ref{fig:green_a} - \ref{fig:green_c}) such as bioswales, porous pavements, and native vegetation restoration,  the program reduces future runoff volumes and contaminant loads entering receiving rivers, thereby mitigating downstream flooding and improving water quality. This case illustrates how targeted land conservation, guided by GIS and hydrologic analysis, can serve as an effective and scalable form of green infrastructure for riverine flood risk reduction. More broadly, U.S. guidance on green infrastructure, climate resilient mitigation, and nature-based solutions provides a consistent framework for integrating these approaches into local hazard mitigation plans, capital planning, and flood risk management programs \cite{ref27}. \\

\section{Soil Stabilization for Flood Prone Subgrades} \label{sec_soil_stabilization}

Soil stabilization is a key strategy for improving the performance and resilience of foundations and pavements in flood-prone floodplains and adjacent low-lying areas. In such environments, subgrades are frequently exposed to high groundwater, cyclic saturation, and soft, fine-grained soils that are vulnerable to loss of strength and excessive deformation under repeated loading. Stabilizers such as lime, cement, and fly ash modify the physicochemical behavior of in situ soils to increase strength and stiffness, reduce plasticity and swell, and enhance durability under wetting, flooding, and freeze–thaw cycles \cite{ref13_diniz2024lime, ref28}. These improvements enable the use of local soils such as subgrade, subbase, and base layers, reducing imported material requirements and supporting more sustainable, climate-resilient infrastructure . \\
\subsection{Lime Stabilization}
Lime stabilization is widely used to improve fine-grained and particularly clayey soils, delivering gains in strength, durability, and resistance to water-related damage. When lime is added to a reactive clay, the pore water pH increases, promoting dissolution of silica and alumina from clay minerals and enabling pozzolanic reactions between calcium from the lime and the solubilized silicates and aluminates. These reactions form calcium silicate hydrates (CSH) and calcium aluminate hydrates (CAH), which progressively cement the soil skeleton, leading to long-term strength gains and improved stability under cyclic saturation and flooding \cite{ref13_diniz2024lime}. Field and laboratory studies show that lime-treated soils can continue to gain strength over decades and maintain stiffness under conditions (soaking, freeze–thaw) that substantially weaken untreated soils \cite{ref29_halim2026stabilization}. \\
Lime treatment provides both short-term modification and long-term stabilization. In the short term, cation exchange and flocculation reduce plasticity, moisture holding capacity, and swell potential, while improving workability and enabling the construction of a stable working platform even in wet conditions \cite{ref30_mukherjee2014selection}. Over longer periods, provided that sufficient lime is present and pH remains above about 10, pozzolanic reactions continue, producing durable cementitious products and increasing resilient modulus and shear strength by an order of magnitude or more in some cases \cite{ref28, ref30_mukherjee2014selection}. National practice reflects this performance, with more than one million metric tons of lime used annually in the United States for soil modification and stabilization beneath roads and other infrastructure. \\
Lime type is critical. Calcitic lime generally contains 90–100\% CaO, whereas dolomitic lime typically contains about 60\% CaO and 40\% MgO. Because magnesium hydroxide is less soluble than calcium hydroxide, MgO reduces the available calcium per unit mass and can limit pozzolanic reactivity unless higher dosages are used. Research on tropical soils demonstrates that calcitic lime produces higher unconfined compressive strength, greater stiffness, and denser, more uniform microstructures than dolomitic lime, which tends to yield more deformable mixtures with non-uniform cementation. Strength increases further as calcitic lime content is raised from about 3\% to 5\%, with continued strength gain observed over extended curing times \cite{ref31_diniz2024lime}. Overall, lime stabilization is particularly effective for high plasticity clays and tropical fine-grained soils, enhancing local soil properties, reducing the need for imported materials, and improving resilience to flood-related damage.

\subsection{Cement Stabilization}
Cement stabilization improves weak or unstable soils by mixing them with Portland cement and water to form a hardened matrix through hydration reactions. Cement hydration produces CSH and CAH gels that bind soil particles and aggregates, significantly increasing load-bearing capacity, reducing plasticity, and improving resistance to moisture, erosion, and repeated traffic loads \cite{ref14, ref32_halsted2006guide}. Owing to rapid early strength gain, cement stabilization is widely applied to road bases, subgrades, and foundations where early opening to traffic and robust structural performance are required. \\
Three standard applications are cement-modified soil (CMS), cement-stabilized subgrade (CSS), and cement-treated base (CTB), differentiated by cement content, degree of bonding, and structural role. Their key characteristics are summarized in Table \ref{tab:cement_stabilization}. Cement stabilization, therefore, provides a flexible toolbox, from modest soil improvement (CMS) to fully structural base layers (CTB). Mix design, soil type, and traffic and environmental conditions determine which application is most appropriate.

\begin{table*}[t]
\small
\centering
\caption{Typical applications of cement stabilization.}
\label{tab:cement_stabilization}
\renewcommand{\arraystretch}{1.2} 
\begin{tabular}{p{3cm} p{4cm} p{4cm} p{4cm} }

\hline
\textbf{Aspect} &
\textbf{CMS – Cement Modified Soil}&
\textbf{CSS – Cement Stabilized Subgrade}&
\textbf{CTB – Cement Treated Base}
\\
\hline
Primary purpose &
Short-term modification: improve workability, drying, and plasticity; stable working platform &
All CMS benefits plus structural support for pavements and foundations &
Structural base layer with high strength and frost resistance \\
\hline
Typical cement content &
$\approx$ 2–4\% of dry soil &
$\approx$ 3–6\% of dry soil &
$\approx$ 3–6\% with selected aggregates or soils \\
\hline
Materials &
In situ fine-grained soils &
In situ fine-grained subgrades (clays or silts) &
Selected granular or on-site aggregates or soils \\
\hline
Material properties &
Unbound or slightly bound; reduced plasticity and moisture susceptibility &
Semi-bound or bound; improved shear and compressive strength; 7-day UCS $\approx$100–300 psi (0.7–2.1 MPa)	&
Fully bound; durable, frost resistant; 7-day UCS $\approx$300–600 psi (2.1–4.1 MPa) \\
\hline
Construction practice &	
Mixed in place; $\geq$ 95\% of maximum dry density; no mellowing period	&
Mixed in place; $\geq$ 95\% of maximum dry density; uniform mixing to full depth	&
Mixed in place or plant mixed; 95–98 \% of maximum dry density; careful curing required \\
\hline
Soil suitability	&
Clayey or silty soils needing workability and moisture control &	
Fine-grained, expansive, or unstable subgrades	&
Coarse aggregates and granular bases under higher traffic \\
\hline
Key benefits	&
Faster construction; uses on-site soils; stable platform	&
Uniform support; non-expensive; can reduce pavement thickness or extend life	&
Thinner pavement sections, high saturated strength, improved frost resistance \\
\hline
Limitations	& 
Non-structural; lower long-term strength &	Primarily subgrade use; typically requires base or surface overlay	&
Requires higher aggregate quality; potential for reflective cracking if not properly designed\\
\hline
\end{tabular}
\end{table*}

\subsection{Fly Ash Stabilization}
Fly ash stabilization modifies soil behavior through pozzolanic and, for certain classes, self-cementitious reactions \cite{ref29_halim2026stabilization}. Class C fly ash contains sufficient calcium and alumina to develop appreciable strength without additional activators and often achieves unconfined compressive strengths exceeding 500 psi under appropriate curing conditions when characterized using practices such as ASTM D5239. When blended with fine-grained soils, fly ash can markedly increase compressive and shear strength, reduce expansive shrink–swell behavior, and improve bearing capacity, especially at addition rates on the order of 8–16\% by dry soil mass. It is also effective at drying wet soils, as the heat of hydration and dilution reduce moisture contents, which is valuable for flood-affected or seasonally saturated subgrades \cite{ref33_senol2006soft}.\\
Studies on Class C fly ash stabilized soft subgrades indicate that strength increases strongly with fly ash content and is sensitive to molding water content and delay between mixing and compaction. Maximum unconfined compressive strength is typically achieved near or slightly above optimum water content and with minimal compaction delay, while longer delays can significantly reduce strength, emphasizing the need for tight construction control. Fly ash stabilization has been shown to reduce swell to less than $\approx$ 0.5\% under moderate confining pressures and to improve resilient modulus sufficiently to allow use of otherwise unsuitable soft clays as subgrade or embankment material. However, performance can be compromised by high sulfate concentrations and organic content, and leaching behavior may require assessment depending on the regulatory context.
\subsection{Comparative Assessment of Lime, Cement, and Fly Ash}
Lime, cement, and fly ash each enhance soil properties through distinct mechanisms, and their relative suitability depends on soil type, target strength, construction schedule, and durability requirements. A concise technical comparison is presented in Table \ref{tab:comparison_stabilization}.

\begin{table*}[t]
\small
\centering
\caption{Comparative characteristics of lime, cement, and fly ash stabilization.}
\label{tab:comparison_stabilization}
\renewcommand{\arraystretch}{1.2} 
\begin{tabular}{p{3cm} p{4cm} p{4cm} p{4cm} }

\hline
\textbf{Parameter} &
\textbf{Lime (Calcitic or Dolomitic)}&
\textbf{Cement (CMS or CSS or CTB)}&
\textbf{Fly Ash (Primarily Class C)}
\\
\hline
Primary mechanism	&
Cation exchange and flocculation; long term pozzolanic reactions at high pH	&
Hydration forming CSH/CAH matrix; rapid binding	&
Self-cementitious (Class C) plus pozzolanic reactions \\
\hline
Typical addition rate (\% soil)	&
$\approx$3–5\% for clays; calcitic often more effective than dolomitic	&
CMS $\approx$2–4\%; CSS $\approx$3–6\%; CTB $\approx$3–6\%	&
$\approx$8–16\% of dry soil (depending on soil and ash class)
\\
\hline
Strength development	&
Moderate early gain; continuous long-term increase over months to decades	& 
Rapid early strength (hours–days); 7-day UCS 100–600 psi depending on application &	Moderate to high; 7-day UCS often 100–500+ psi; sensitive to moisture and delay\\
\hline
Best soil types	&
High plasticity clays; tropical fine grained soils	&
Fine-grained subgrades (CMS/CSS) and granular bases (CTB)	&
Expansive or plastic clays; wet soft subgrades\\
\hline
Plasticity / swell control	&
Large reduction in PI and swell; improved workability and moisture memory	&
Reduced PI and shrink–swell for CMS/CSS; non-expansive subgrade when well designed &	Strong swell control ($<\approx$0.5\% swell in some tests); PI less critical\\
\hline
Durability under flooding and cycles &	
Good long-term durability; resilient to soaking, freeze–thaw, and repeated wetting if the mix is properly designed	&
CTB and well-designed CSS provide high saturated strength and frost resistance	&
Good durability if sulfates/organics controlled; may require leachate assessment\\
\hline
Construction sensitivity	&
Requires proper mixing, compaction, and curing; mellowing period for heavy clays &
Requires density control and curing; CTB may need crack mitigation	&
Sensitive to moisture content and compaction delay; sulfate and organics can limit use\\
\hline
Sustainability and resource use	&
Uses widely available lime; enables reuse of in situ soils and reduces imported materials	&
Enables use of on-site soils and thinner sections; cement production has higher embodied CO2	&
Utilizes industrial byproducts; can reduce costs and environmental impact where locally available\\
\hline
Typical application	& 
High plasticity clays, tropical soils, and flood-prone subgrades needing long-term resilience	&
Rapid, structural stabilization for high traffic pavements and foundations (especially CTB)	&
Eco efficient stabilization of soft or expansive soils in low  to moderate traffic applications\\
\hline
\end{tabular}
\end{table*}

In combination, these stabilizers provide a versatile toolkit for designing resilient, cost-effective foundations and pavements in flood-affected environments. Lime is typically favored for high plasticity clays and long-term resilience, cement for applications requiring rapid, high structural strength, and fly ash for eco-efficient stabilization of expansive, wet, or soft soils, particularly where its by-product nature aligns with sustainability goals. Blended systems (such as lime–fly ash) can further optimize performance and should be supported by site-specific laboratory mix design and durability testing.


\section{Resiliency Improvements for Existing Substation and Related Upgrade} \label{sec_rexisting_stations}
Existing transmission and distribution substations face growing flood risk as hydrometeorological extremes intensify under climate change. Approximately 9–14\% of U.S. transmission and distribution, predominantly substation assets, are located within FEMA 100-year (1\% AEP) floodplains, with roughly 10\% (about 8,000 assets) in mapped 500-year (0.2\% AEP) zones \cite{ref4_tufail2026assessing}. A further 2–5\% of facilities, approximately 1,600–4,000 sites, lie in coastal surge V zones where combined riverine and coastal flooding can cause deep, high velocity inundation \cite{ref16_prime2018protecting}. These exposures have translated into cascading grid failures, outages affecting thousands to millions of customers, and restoration delays that contribute to substantial direct and indirect economic losses in the United States each year. Table \ref{tab:exposure} summarizes representative estimates of substation exposure by flood zone.

\begin{table*}[t]
\small
\centering
\caption{Representative exposure of U.S. substations to flood zones}
\label{tab:exposure}
\renewcommand{\arraystretch}{1.2} 
\begin{tabular}{p{3cm} p{2cm} p{2cm} p{2cm} p{3cm} }

\hline
\textbf{Flood zone} &
\textbf{Description}&
\textbf{Approx. share of substations}&
\textbf{Estimated number} &
\textbf{Primary data sources}
\\
\hline
100 year (1\% AEP) &
FEMA Zones A / AE	&
9–14\%	&
Varies by source	&
FEMA DFIRM, HIFLD infrastructure data\\
\hline

500 year (0.2\% AEP)	&
FEMA 0.2\% AEP floodplain &
$\approx$10\%	&
$\approx$8,000	&
FEMA / NOAA overlays, national studies\\
\hline
Coastal surge (V zones)	&
High energy coastal flooding	&
2–5\%	&
1,600–4,000 &	
ASCE 24 coastal criteria, coastal mapping\\
\hline

\end{tabular}
\end{table*}

Flooded substations can precipitate region-wide instability because they concentrate high voltage transformation and switching functions. Historical events and recent spatial–exposure analyses show that outages at a relatively small set of flood-prone nodes can propagate across networks, extend restoration times, and generate large annualized losses when climate-amplified storms affect clustered assets. As a result, there is a pressing need for both immediate and long-term hardening measures at existing substations, not just at new (greenfield) sites.\\
A two-track strategy is usually recommended, with short-term measures focusing on rapidly deployable, non-intrusive defenses that can be implemented with zero or minimal outages, such as temporary flood barriers, rapidly deployable pumps, and localized earth berms \cite{ref16_prime2018protecting}. Long-term measures include structural and geotechnical upgrades, such as raising equipment elevations, installing permanent flood barriers, enhancing yard drainage, retrofitting foundations for scour resistance, and deploying real-time sensing and control systems to support anticipatory operation during floods \cite{ref17}. To be practical, these measures must be organized into a phased program that aligns with regulatory frameworks such as FEMA Risk MAP, ASCE 24 Class IV, and NERC CIP 014, and leverages lessons from recent utility case studies. \\
\subsection{Phased Upgrade Framework for Existing Substations}
\begin{figure*}[!htbp]
  \centering
  \includegraphics[width=0.60\textwidth]{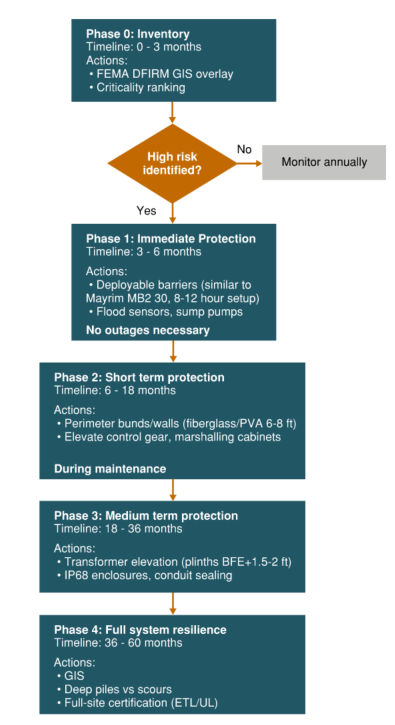}
    \caption{Proposed conceptual flowchart for existing substation flood resilience upgrade}
    \label{fig2}
\end{figure*}

A hybrid, multi-phase approach can provide a structured path to 500-year (0.2\% AEP) resilience for existing substations over a 3–5 year period while minimizing downtime. Figure \ref{fig2} conceptually illustrates the sequence from initial assessment through full system resilience.

\subsection*{Phase 0 – Inventory and Assessment (0–3 months)}
Phase 0 establishes the analytical foundation for all subsequent interventions. The primary objective is to identify flood-vulnerable substations and prioritize them based on exposure and criticality. This is achieved by overlaying substation locations with FEMA DFIRM floodplain data and other flood hazard layers in a GIS environment, enabling rapid identification of assets in 100-year, 500-year, and coastal flood zones \cite{ref15}. Hydrologic and hydraulic modeling using tools such as HEC RAS is then used to estimate flood depths, velocities, and durations under 500-year (and selected intermediate) scenarios at each high-risk site \cite{ref27}.\\
Criticality ranking is performed based on load served, role in network topology, restoration time, and the potential socio-economic impacts of prolonged outages. The key deliverables of Phase 0 include a prioritized list of high-risk substations and baseline risk metrics (such as qualitative FEMA Risk Index scores) that can guide investment planning \cite{ref16_prime2018protecting}.\\

\subsection*{Phase 1 – Immediate Protection (3–6 months, zero outage)}
Phase 1 focuses on rapid deployment of non invasive protective measures that can be installed with little or no interruption to substation operations. The objective is to reduce near-term flood risk while more comprehensive upgrades are designed and scheduled. Typical interventions include modular, portable flood barrier systems, additional sump pumps and non-return valves for critical low points, and flood level sensors integrated with SCADA and alarm systems. Case studies show that utilities have successfully deployed demountable aluminum barrier systems and mobile pumping units at high-risk sites to prevent or limit inundation during forecast events, in some cases without requiring extended outages.\\
Equipment used in this phase is typically specified for high ingress protection ratings (e.g., IP67 or better) and for continuous operation under multi-day submersion scenarios, reflecting the need for robustness during prolonged events. Because Phase 1 measures are largely external and reversible, they provide an important bridge from current conditions to full structural hardening.\\

\subsection*{Phase 2 – Perimeter Defense and Drainage (6–18 months)}
Phase 2 establishes a primary flood exclusion zone around the substation through more permanent perimeter defenses and improved site drainage. This may involve the installation of fixed or semi-permanent floodwalls or berms, often constructed from fiberglass, PVC, or reinforced concrete with integrated flood gates and vehicular access points. In several documented programs, utilities have implemented 6–8 ft high perimeter walls at multiple high-risk sites to contain fluvial and pluvial flooding while maintaining safe crew access. \\
At the same time, yard grading and drainage improvements are designed and verified using hydraulic modeling to ensure that 500-year design flows can be conveyed without passing protective structures or creating hazardous internal ponding \cite{ref4_tufail2026assessing}. Where feasible, these works are coordinated with scheduled maintenance outages to elevate control gear and marshaling kiosks, relocate critical low-voltage equipment above anticipated flood levels, and integrate green infrastructure features such as bioswales and infiltration areas at the site boundary. Proper anchoring and foundation design for perimeter structures is essential to withstand combined hydrostatic, hydrodynamic, and wind loads, with embedment depths frequently extending several meters to ensure stability. \\

\subsection*{Phase 3 – Core Asset Protection (18–36 months)}
Phase 3 targets the highest value and most impact-sensitive equipment within the substation, including power transformers, high voltage switchgear, and control buildings. The primary objective is to ensure that these assets are protected to or above the required design flood elevation, often defined as the base flood elevation (BFE) plus a freeboard margin of 1.5–2 ft in line with ASCE 24 Class IV guidance for essential facilities. Measures include elevating transformers on reinforced concrete plinths or platforms, upgrading or replacing switchgear with flood resistant (e.g., IP68 rated) enclosures, and sealing cable trenches and duct banks using approved sealing systems to prevent water ingress.\\
Because these interventions can affect core electrical clearances and network configuration, they are typically implemented during planned outages coordinated with regional system operators, sometimes using mobile or temporary units to maintain service continuity. Experience from utility projects shows that, with careful planning and temporary redundancy, elevation and enclosure upgrades can often be completed with disruptions limited to a few days per asset.\\

\subsection*{Phase 4 – Full System Resilience (36–60 months)}
Phase 4 aims to achieve comprehensive 500 year design performance and formalize the substation’s status as a resilient, climate ready asset. This includes completing the transition to flood-resistant GIS switchgear where appropriate, retrofitting foundations with deep piles or helical anchors to resist scour and erosion in accordance with relevant bridge scour and flood design standards, and integrating microgrid tie-ins that allow controlled islanding of critical loads during grid-wide disturbances. Energy storage systems, UPS units, and control equipment are elevated above design flood levels, and all major components are evaluated against relevant IEEE, NERC CIP 014, and certification standards for flood resilience\cite{ref16_prime2018protecting}. \\
At this stage, the substation’s physical, geotechnical, and green infrastructure measures are integrated into a unified monitoring and control system, providing real-time situational awareness and enabling adaptive responses during evolving flood conditions \cite{ref27}. Third-party testing and certification may be pursued to demonstrate compliance with specified flood-resilience criteria, thereby supporting regulatory approval and stakeholder confidence. \\
Table \ref{tab:phase_compare} provides a concise overview of the phased upgrade strategy. This phased framework demonstrates how existing substations can transition from high exposure and reactive recovery toward proactive, integrated resilience within a practical planning horizon.

\begin{table*}[t]
\small
\centering
\caption{Summary of phased existing-substation flood-resilience upgrade}
\label{tab:phase_compare}
\renewcommand{\arraystretch}{1.2} 
\begin{tabular}{p{1.5cm} p{2cm} p{2cm} p{4cm} p{4cm} }

\hline
\textbf{Phase} &
\textbf{Timeframe}&
\textbf{Primary objective}&
\textbf{Key actions} &
\textbf{Typical constraints/notes}
\\
\hline
0	&
0–3 months	&
Identify and prioritize high risk sites	&
GIS–FEMA overlay; HEC RAS modeling; criticality ranking	&
Desktop and limited field verification\\
\hline
1	&
3–6 months	&
Immediate, zero outage protection	&
Deployable barriers; pumps; flood sensors and alarms	&
No or minimal outages; reversible measures\\
\hline
2	& 
6–18 months	&
Perimeter defense and drainage	&
Perimeter walls/berms; gate structures; drainage regrading and modeling	&
Coordinate with maintenance; access maintained\\
\hline
3	&
18–36 months	&
Core asset protection	&
Elevate transformers; upgrade switchgear; seal ducts; temporary bypass	&
Planned outages; detailed electrical integration\\
\hline
4	&
36–60 months	&
Full system resilience and certification &	GIS switchgear; scour-resistant foundations; microgrid tie-ins; standards compliance	 &
Highest capital intensity; long-term benefits\\
\hline

\end{tabular}
\end{table*}

\subsection {Non-Stationarity and Multi-Scalar Resilience in Substation Stormwater Design}
Hydroclimatic extremes such as intense rainfall, floods, and droughts no longer exhibit stable statistical behavior, largely due to anthropogenic climate change. Instead of following time-invariant probability distributions, key variables now show shifting means, variances, and tail behaviors, which undermines the long-standing assumption that historical records provide a reliable basis for infrastructure design \cite{ref18_markolf2021re, ref34}. Traditional stormwater and flood control criteria, for example, “100-year storms” derived from historical intensity–duration–frequency (IDF) curves, are therefore increasingly inconsistent with the conditions electrical substations will face over their design life. In practice, this mismatch can manifest as drainage and yard grading systems that are routinely exceeded, equipment elevations that are no longer adequate, and misallocated investments in defenses tuned to obsolete hazards. These shortcomings motivate a shift from purely stationary, design storm-based methods toward multi-scalar resilience strategies that explicitly address non-stationarity and the stormwater vulnerabilities of substations \cite{ref19_kim2022leveraging}.

\subsubsection {Non-Stationarity in Hydroclimatic Design}
In statistical terms, a process is stationary if its joint probability distribution is invariant under shifts in time, meaning that its moments and dependence structures do not change \cite{ref35_kaplan1981quantitative}. Under non-stationary conditions driven by external forcings, most prominently greenhouse gas-induced warming, this invariance fails, and the statistical properties of precipitation, temperature, and related hydrologic variables become time dependent \cite{ref36_milly2008stationarity}. Several features are especially relevant for stormwater and substation design:
\begin{itemize}
    \item \textbf{Trend shifts in central tendency and dispersion} Warming increases atmospheric moisture capacity and modifies circulation patterns, leading to systematic changes in mean precipitation and variability in many regions; this can increase both average rainfall and the variability of wet and dry spells \cite{ref36_milly2008stationarity}.
    \item \textbf{Altered tail behavior of extremes} Return levels associated with events historically labeled as “100 year” or rarer storms have intensified, while effective return periods have shortened in numerous basins \cite{ref37_cheng2014nonstationary}.
    \item \textbf{Anthropogenic attribution} Detection and attribution analyses show that many observed changes in heavy precipitation and flood frequency cannot be explained by natural variability alone and are linked to anthropogenic forcing, implying that unadjusted historical data systematically understate future risks \cite{ref34}.
    \item \textbf{Coupling with land use and exposure changes} Urbanization increased impervious area, and expansion into floodplains elevates runoff generation and concentrates assets in hazardous locations, further amplifying effective non-stationarity in risk \cite{ref18_markolf2021re}.
\end{itemize}

Table 6 summarizes these drivers and their implications.

\begin{table*}[t]
\small
\centering
\caption{Drivers of hydroclimatic non-stationarity and design implications}
\label{tab:phase_compare}
\renewcommand{\arraystretch}{1.2} 
\begin{tabular}{p{3.5cm} p{4cm} p{4cm}}

\hline
\textbf{Driver} &
\textbf{Description/ examples}&
\textbf{Implications for stormwater and substations}
\\
\hline
Climate-induced trend shifts &	
Systematic changes in mean rainfall and temperature	&
Historical averages lose predictive value\\
\hline
Intensified extremes &
Higher peak intensities; shorter effective return periods	&
A higher degree of vulnerability from legacy “100-year” storm designs\\
\hline
Anthropogenic attribution	&
Forced climate change beyond natural variability	&
Requires climate-informed, forward-looking design\\
\hline
Land use and exposure changes	& Urbanization, development in floodplains &	Higher runoff, asset clustering, larger losses\\
\hline
Socioeconomic dynamics	&
Population growth, critical infrastructure dependence	&
Greater consequence side of risk\\
\hline
\end{tabular}
\end{table*}

\subsubsection{Limitations of Stationary Design Storm Criteria}
Under non-stationary conditions, three core limitations of the stationary design storm criteria become evident, especially for stormwater systems serving substations. These are discussed in the following sections:
\begin{enumerate}
    \item \textbf{Data invalidity} Frequency-based design tools assume that historical extremes reliably represent future behavior. As climate-driven shifts alter both mean conditions and tail behavior, IDF curves and return periods estimated under stationarity can substantially mischaracterize future storm intensities. Studies have shown that climate-informed projections often indicate increased design storm magnitudes by mid-century and widespread exceedance of existing drainage capacities, with accompanying increases in maintenance and rehabilitation costs for pavements and associated infrastructure \cite{ref38_gudipudi2017impact}. For substations, this translates into more frequent yard ponding, overtopping of perimeter channels, and flooding of equipment thought to be above design levels.
    
    \item \textbf{Risk triplet gaps} In the Kaplan–Garrick formulation, risk is described as a set of triplets comprising hazard, probability, and consequences [\cite{ref35_kaplan1981quantitative}. Conventional design storms address hazard and probability but often treat consequences implicitly, for example, by assigning the same return period design to assets with very different criticalities. This can leave highly critical substations—serving hospitals, data centers, or large urban loads—underprotected relative to their potential impact if flooded \cite{ref19_kim2022leveraging}.
    
    \item \textbf{Multi-hazard and cascading effects} Single-hazard design criteria typically neglect compound events and interdependencies between systems. In reality, substation flood risk may be heightened by concurrent power outages at pumps, blocked drainage from debris, or restricted access routes, all of which can prolong inundation or complicate recovery \cite{ref18_markolf2021re}. Ignoring these interactions can result in designs that appear adequate under isolated assumptions but fail under realistic multi-hazard scenarios.
\end{enumerate}
   These limitations lead directly to the conclusion that stationary, single-hazard design storm approaches are insufficient for substations exposed to non-stationary hydroclimatic risks. A more promising pathway is to embed stormwater design within a multi-scalar resilience framework that reflects component-, system-, and portfolio-scale perspectives.

\subsubsection {Multi-Scalar Resilience Strategies for Substation Stormwater Design}
Electrical substations aggregate transformers, switchgear, control buildings, and cable systems, often in low-lying or constrained sites chosen for network topology and access. Stormwater vulnerabilities include inadequate grading and drainage, limited onsite storage, insufficient freeboard for equipment, and a lack of controlled overflow paths. Floodwaters can cause short circuits, insulation breakdown, oil releases, mechanical damage to cooling systems, and extended outages that propagate through the grid. Under non-stationary conditions, the frequency and severity of such events can increase, making it essential to design for resilience across multiple scales: component, system, and portfolio scales. These measures create buffer capacity at the multiple hierarchical levels, as elaborated in the following section: 
\begin{enumerate}
    \item \textbf{Component scale resilience: Hardening individual assets}
    At the component scale, the objective is to reduce the probability and consequences of failure for individual assets subjected to stormwater loads. Representative strategies include:
    \begin{itemize}
    \item Elevation of critical equipment -  Raising transformers, relay panels, control cabinets, and other key components above non-stationary design flood elevations (e.g., 500-year level plus freeboard) using plinths or platforms that can be adjusted or augmented over time.
    \item Flood-resistant enclosures and penetrations - Specifying switchgear enclosures with appropriate ingress protection ratings, sealing cable trenches and duct banks, and using water-tight building penetrations to reduce water ingress and electrical faults during shallow flooding.
    \item Localized drainage and pumping -  Installing sumps and pumps with backup power and non-return valves near critical equipment to manage localized ponding when site drainage is overloaded.
    \item Sensing and automated triggers -  Deploying water level, conductivity, and structural sensors connected to SCADA and protection systems to enable early warning, preemptive switching, or controlled shutdown when thresholds are approached.
\end{itemize}
\item \textbf{System scale resilience: Safe to fail substation precincts} At the system scale, resilience focuses on the substation precinct as an integrated hydrologic and electrical system. Recognizing that extreme events may exceed nominal design thresholds, safe-to-fail design aims to manage exceedance in a controlled manner rather than preventing it at all costs. Key elements include:
\begin{itemize}
    \item Zoned yard configurations -  Organizing the yard into sacrificial zones designed for temporary storage or flow conveyance and heavily protected core zones containing critical equipment. Sacrificial zones may include gravel sumps, depressed areas, or access roads that can safely pond water without damaging equipment.
    \item Redundant drainage and bypass paths - Designing swales, channels, and overflow structures to direct excess runoff toward retention or infiltration areas, or safely off-site, rather than across critical equipment pads.
    \item Hybrid gray–green stormwater systems - Combining engineered measures such as ACB-lined channels, berms, and floodwalls with bioswales, rain gardens, and permeable pavements to attenuate peak flows, enhance infiltration, and improve water quality.
    \item Operational exceedance protocols - Establishing procedures for barrier deployment, pump operation, and switching configurations during forecast or real-time exceedance events, ensuring that hydrologic pathways and electrical operations remain consistent with safe-to-fail objectives.
\end{itemize}
Hydrologic hydraulic modeling that reflects non-stationary scenarios, yard topography, and internal drainage structures is essential to designing and evaluating such precinct-scale configurations.
\item \textbf{Portfolio scale resilience: Network level planning}
 At the portfolio scale, resilience is addressed across fleets of substations and associated infrastructure. The focus is on optimizing where and how to invest in stormwater and flood resilience measures under deep uncertainty about future climate and development trajectories. Core components include:
\begin{itemize}
    \item Risk-based prioritization - Applying consistent exposure, vulnerability, and criticality metrics across all substations to identify high-risk sites and prioritize interventions based on risk reduction per unit cost.
    \item Scenario-based and robust decision frameworks - Evaluating portfolios of adaptation options, combinations of component scale hardening and system scale reconfiguration across multiple climate and land use scenarios to identify strategies that perform satisfactorily across a wide range of futures rather than optimizing for a single forecast.
    \item Digital twins and integrated modeling - Developing coupled hydrologic-hydraulic power system models to simulate how flooding at one or more substations affects network performance, enabling virtual testing of stormwater interventions and operational strategies before deployment.
    \item Mutual aid and recovery planning. Establishing mutual aid agreements, pre-positioned mobile equipment, and coordinated restoration strategies that reduce recovery time when flooding occurs.
\end {itemize}

Multi-scalar resilience strategies, as outlined in the preceding section and as summarized in Figure \ref{fig:Multi_scalar_resilience} for substation stormwater design, should be developed in a staged, hierarchical manner - progressing from component-scale hardening, to system-scale safe-to-fail design, and ultimately to network-level portfolio optimization, to ensure that resilience is built incrementally while maintaining coherence across spatial and operational scales. By explicitly linking non-stationary hydroclimatic behavior to the limitations of stationary design storm criteria and then to a structured, multi-scalar resilience framework, this section offers a conceptual basis for reconsidering stormwater design criteria and flood resilience at substations.

\begin{figure*}[!htbp]
  \centering
  \includegraphics[width=0.95\textwidth]{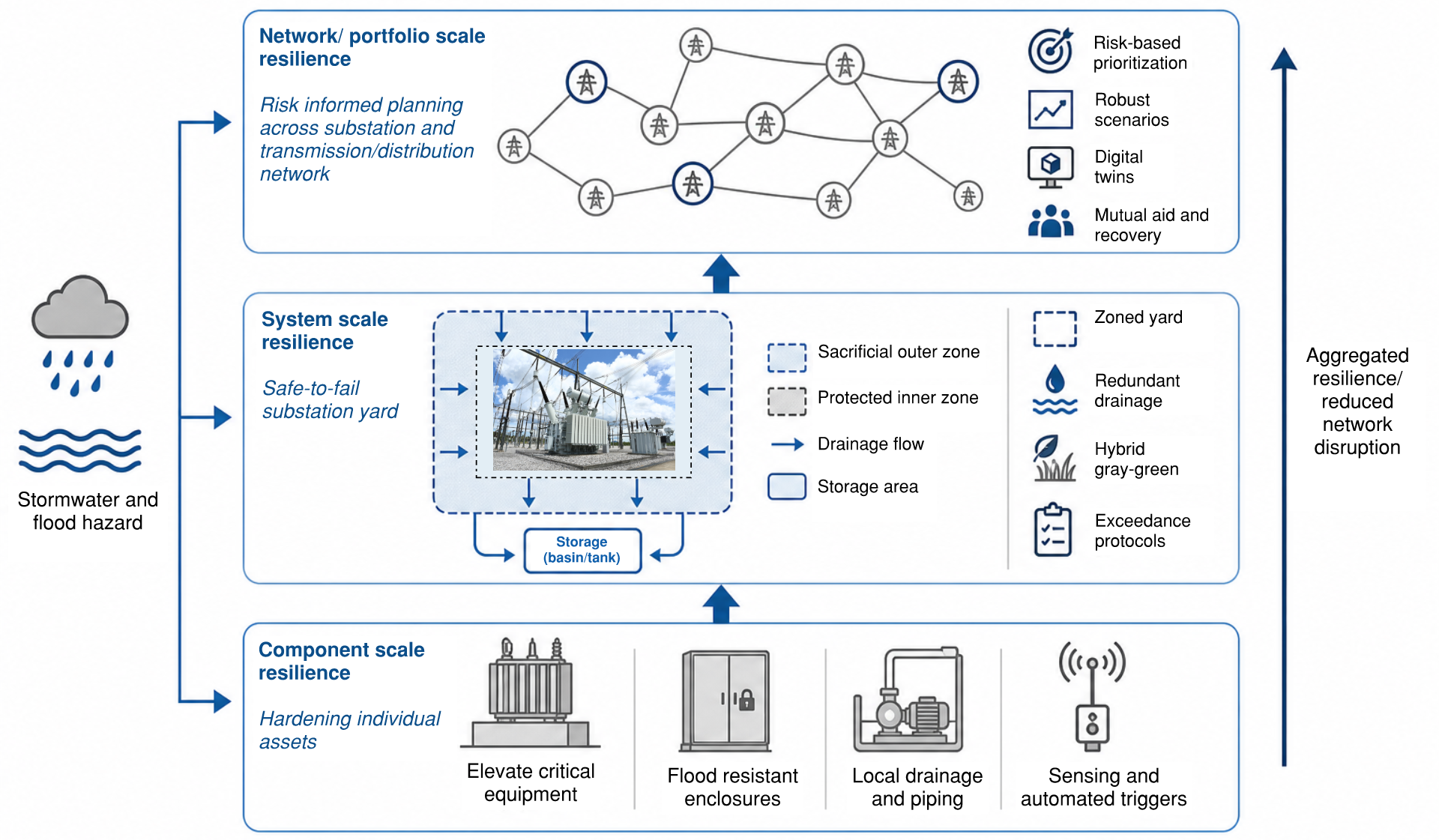}
    \caption{Three‑tier multi‑scalar resilience workflow for substation stormwater design: component, precinct, and network scales}
    \label{fig:Multi_scalar_resilience}
\end{figure*}

\item \textbf{Application of explainable AI (XAI) in portfolio-scale resilience: Leveraging AI to tie it all} As utilities increasingly deploy machine learning to forecast and manage flood risk across substations, feeders, and distributed energy assets, the opacity of conventional "black-box" AI models has become a critical adoption barrier. Explainable AI (XAI) directly addresses this gap: by attributing model outputs to specific input features, methods such as SHAP (Shapley Additive Explanations) and LIME (Local Interpretable Model-Agnostic Explanations) reveal which variables (rainfall intensity, soil moisture, topography, load prioritization) drive a given prediction, transforming a probabilistic output into an actionable, auditable insight. This is the core leverage of XAI for critical infrastructure: it converts predictive power into institutional trust, regulatory defensibility, and operational accountability, all of which are prerequisites for deploying AI in decisions that affect public safety and grid reliability. For utilities managing a geographically distributed portfolio of substations, feeders, and distributed energy resources, XAI shifts flood resilience from a reactive, single-asset exercise into a coordinated, explainable, portfolio-wide decision process. The research and deployment in the field of XAI based portfolio-scale resilience has started to gain traction; for example, \cite{wei2024using} in their research outlines methods to understand the nonlinear relationship between the Three Gorges Dam and downstream flood.
The following four functions illustrate the leverage XAI provides.
\begin{itemize}
    \item Risk assessment and prediction with transparent explanations: Portfolio-level resilience begins with predicting which assets are most vulnerable to an approaching flood event, but a numeric risk score alone is insufficient for operational decision-making. Feature-attribution techniques such as SHAP and LIME \cite{hSHAP_2025} quantify each input variable's contribution to a given asset's predicted flood risk, allowing engineers to trace a "high vulnerability" flag for a specific substation back to concrete drivers such as low elevation, proximity to a stream, or antecedent soil saturation. SHAP, LIME, and attention mechanisms are the most widely applied XAI techniques in flood forecasting, hazard mapping, and decision-support systems, consistently improving stakeholder engagement with model outputs. Applied to a portfolio of assets, this means risk rankings can be defended asset-by-asset, rather than treated as an unexplained black-box score, which is essential when prioritizing limited hardening budgets across a service territory.
    \item Resource allocation: Once vulnerable assets are ranked with explained justifications, XAI supports the allocation of finite hardening and response resources, such as mobile generators, sandbags, crew deployments, or spare transformers, to the locations where marginal risk reduction is greatest.
    \item Emergency power management and coordination: During an active flood event, operators must make rapid decisions such as load shedding, microgrid islanding, or feeder reconfiguration, often through reinforcement-learning-based control agents whose recommendations must be trusted in real time. Work applying Deep-SHAP to deep reinforcement learning models for power system emergency control shows that feature-level attributions can clarify why an RL agent recommends specific actions, giving human operators the interpretable basis needed to accept or override automated coordination decisions. 
    \item Post-flood analysis: After a flood recedes, XAI may support root-cause analysis of what actually happened across the portfolio versus what was predicted, closing the loop for model refinement and infrastructure hardening. Attention mechanisms and uncertainty-aware explanation methods have been used to trace compound flooding effects (e.g., rainstorm and tidal interaction) back to specific driving variables, helping engineers understand why certain assets failed or performed better than predicted. A typical workflow is shown in Figure \ref{fig:XAI_chart__post_flood_xai_lstm}, and consists of building the post-flood analysis prediction model as step 1, using XAI methods to calculate feature importance as step 2, and analyzing and ranking the key drivers, as step 3. 

\begin{figure*}[!htbp]
  \centering
  \includegraphics[width=0.95\textwidth]{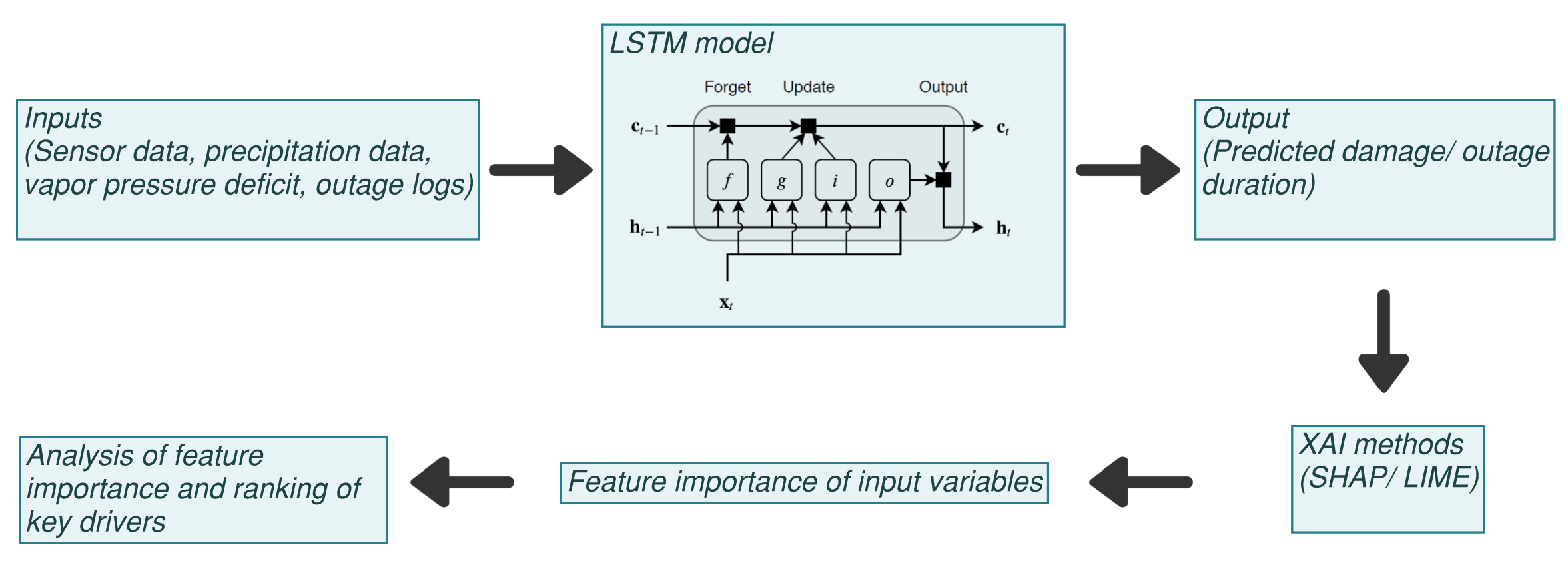}
    \caption{Flowchart showing XAI based post-flood analysis}
    \label{fig:XAI_chart__post_flood_xai_lstm}
\end{figure*}

\end{itemize}
\end{enumerate}

\section{Conclusion} \label{sec_conclusion}
The work presented in this manuscript shows how flood‑exposed substations can be re‑imagined as resilient infrastructure by treating erosion control, geotechnical stabilization, green infrastructure, and retrofit planning as parts of the same engineered system. By combining permeable ACB revetments, lime/cement/fly‑ash stabilization, and stormwater‑focused green infrastructure, the framework moves existing brownfield yards toward non‑stationary 500‑year (0.2\% AEP) performance while reducing life‑cycle costs and aligning with NERC CIP‑014, ASCE 24 Class IV, FEMA Risk MAP, and LEED objectives. The phased roadmap proposed in this work offers utilities a practical way to build resilience in small, manageable steps rather than through disruptive, single‑stage rebuilds.
Equally important, the manuscript reframes substation design as a portfolio‑scale resilience challenge under non‑stationary climate and land‑use conditions, rather than a site‑by‑site compliance exercise. It connects hydrologic‑hydraulic modeling, GIS exposure mapping, and explainable AI tools such as SHAP and LIME to risk‑based prioritization, so that investment decisions across fleets of substations can be transparent, traceable, and defensible.

\section*{Generative AI Usage Statement}
During the preparation of this work, the author(s) used Claude (Anthropic) to assist with sentence formatting and grammar checking. After using this tool, the author(s) reviewed and edited the content as needed and took full responsibility for the content of the published article.

\section*{Author Statement}
\textbf{CS:} Conceptualization, Methodology, Formal analysis, Writing - Original Draft, Writing - Review and Editing, Visualization. \textbf{SG:} Conceptualization, Methodology, Formal analysis, Supervision Writing - Original Draft, Writing - Review and Editing.

\section*{Funding}
This research received no external funding. This work was independently conducted and self-funded by the authors.

\section*{Declaration}
The authors declare that they have no known competing financial interests that could have appeared to influence the work reported in this paper. \\
The views, thoughts, opinions, and conclusions made in this material are solely those of the authors and don’t necessarily reflect the views of the authors' employer, organization, committee, or other group or individual. \\

\textbf{Ethics, Consent to Participate, and Consent to Publish declarations: not applicable.}\\

\textbf{Clinical trial number: not applicable.}\\
\inConf{
\bibliographystyle{IEEEtran}
\bibliography{references}
}
\inArxiv{
\bibliographystyle{tmlr}
\bibliography{references}
}


\end{document}